\documentclass[aps,prb,showpacs,twocolumn,groupedaddress]{revtex4}
\usepackage{graphicx}% Include figure files
\usepackage{dcolumn}% Align table columns on decimal point
\usepackage{bm}% bold math
\begin{document}

%\preprint{}

\title{Quantum interference of electrons in $\text{Nb}_{5-\delta}\text{Te}_4$
single
%%%crustals
crystals}


\author{A.~Stolovits}
\email[Electronic address: ]{yas@fi.tartu.ee}
\author{A.~Sherman}
%\thanks{}
\affiliation{Tartu \"Ulikooli F\"u\"usika Instituut, Riia 142, EE-51014 Tartu,
Estonia}
\author{R.~K.~Kremer}
\author{Hj.~Mattausch}
\author{H.~Okudera}
\author{X-M.~Ren}
\altaffiliation[Present address: ]{Research Institute for Electronics Science,
Hokkaido University, Sapporo 060-0812, Japan.}
\author{A.~Simon}
\affiliation{Max-Planck-Institut f\"ur Festk\"orperforschung,
Heisenbergstrasse 1, D-70569 Stuttgart, Germany}
\author{J.~R.~O'Brien}
\affiliation{Quantum Design, 6325 Lusk Boulevard, San Diego, CA 92121}

\date{\today}

\begin{abstract}
The compound $\text{Nb}_{5-\delta}\text{Te}_4$ ($\delta=0.23$)
with quasi-one-dimensional crystal structure undergoes a
transition to superconductivity at $T_c$=0.6--0.9~K. Its
electronic transport properties in the normal state are studied in
the temperature range 1.3--270~K and in magnetic fields up to
11~T. The temperature variation of the resistivity is weak
($<2\%$) in the investigated temperature range. Nonmonotonic
behavior of the resistivity is observed which is characterized by
two local maxima at $T\sim$2~K and $\sim$30~K. The temperature
dependence of the resistivity is interpreted as an interplay of
weak localization, weak antilocalization, and electron-electron
interaction effects in the diffusion and the Cooper channel. The
temperature dependence of the dephasing time $\tau_\varphi$
extracted from the magnetoresistance data is determined by the
electron-phonon interaction. The saturation of $\tau_\varphi$ in
the low-temperature limit correlates with $T_c$ of the individual
crystal and is ascribed to the scattering on magnetic impurities.
\end{abstract}

%\keywords{}

\maketitle

\section{INTRODUCTION}

In disordered metals the coherence of the conduction electrons may
extend over large distances and exceed %%%
the mean free path by several orders of magnitude. This %%%contributes
large scale coherence manifests itself in %%% into
interference effects such as  %%%
weak localization, interference corrections to electron-electron interaction
and various mesoscopic phenomena.\cite{alt85,imry_book} In general, the
temperature dependence of the dephasing time $\tau_\varphi$ is governed by
electron-electron and electron-phonon interaction. In disagreement with the
standard theory of
electron dephasing,\cite{alt82} %%%
a saturation of $\tau_\varphi$ in the low-temperature limit has been
observed in numerous experiments. %%%Beside the statements
It was suggested that this saturation is universal and reflects
fundamental properties of disordered conductors.\cite{moh97,lin01} %%%other
%%%
Various mechanisms for the saturation behavior have been discussed, including
effects of magnetic impurities, tunneling two-level systems, electron
heating, and separated superconducting grains\cite{lar04} %%%For a recent review
(see Refs.~\onlinecite{bird,esteve}).

Recently the %%%disorder quality
character of the disorder %%%???
and its influence on $\tau_\varphi$ %%%
have become an important issue of research. Weak localization studies of
differently prepared PdAg films show that the microscopic structure of disorder
determines the interaction of the electrons with the phonons.\cite{lin02b}
%%%
The nature of the scattering potential plays a crucial role in the
Sergeev-Mitin electron-phonon interaction theory.\cite{ser00} While in dirty
systems the scattering on vibrating impurities results %%%
in a $T^4$ dependence of the electron-phonon scattering rate
$\tau_{ep}^{-1}$, scattering on %%%
a static potential leads to
$\tau_{ep}^{-1} \sim T^2$. Disorder %%%
also influences %%%
the %%%low-temperature
saturation value $\tau_0$ of the dephasing time. %%%
For example, Lin {\it et al.} found that annealing of moderately
disordered
three-dimensional polycrystalline metals %%%
raises $\tau_0$.\cite{lin02} %%%
This result can be explained in terms of two-level systems
associated with the grain boundaries. However, annealing effects
have not been observed in strongly disordered samples.\cite{lin02}
To further investigate disorder effects, systems in which the
crystallinity can be tuned in a broad range are desirable. Single
crystalline systems with disorder, e.~g. emerging from vacancies,
which exhibit quantum interference effects %%%,
are potential candidates for such studies.

\begin{figure}
\includegraphics[width=8cm]{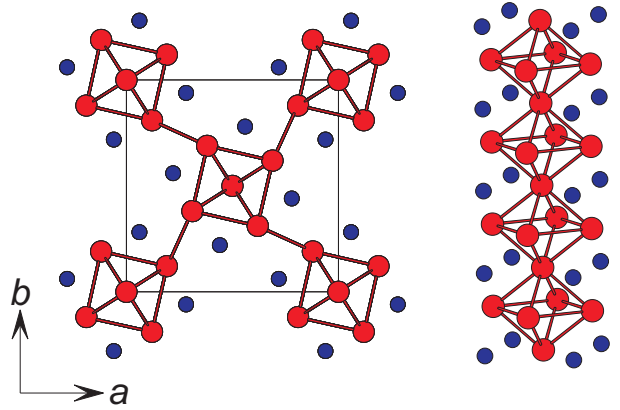}
\caption{\label{structure}Projection of the Nb$_5$Te$_4$ structure along the
$c$-axis of the tetragonal unit cell (left) and perspective view of its single
chain of Nb$_6$ octahedra (right). Large circles indicate Nb atoms, small
circles  represent Te atoms.}
\end{figure}

In this paper we present electron transport studies on single
crystals of $\text{Nb}_{5-\delta}\text{Te}_4$.
$\text{Nb}_{5-\delta}\text{Te}_4$ belongs to the growing family of
compounds crystallizing with the tetragonal
$\text{Ti}_{5}\text{Te}_4$ structure-type (space group
$I4/m$).\cite{set63} For a list of presently known compounds with
this structure-type
see %%%in
Ref.~\onlinecite{he}. The basic %%%
elements of the structure %%%is
are compressed $\text{Ti}_{6}\text{Te}_8$ %%%
clusters which condense to
form infinite $\text{Ti}_{5}\text{Te}_4$ chains. These chains are linked
through Ti-Te and Ti-Ti contacts (see Fig.~\ref{structure}).
$\text{Ti}_{5}\text{Te}_4$,
$\text{Nb}_{5}\text{Te}_4$, $\text{Mo}_{5}\text{As}_4$,
$\text{Nb}_{5}\text{Sb}_4$, and $\text{Nb}_{5}\text{Se}_2\text{S}_2$ are
reported to be metals.\cite{set65,dron90,ts} Among them
$\text{Nb}_{5}\text{Sb}_4$ and $\text{Nb}_{5}\text{Se}_2\text{S}_2$ are
superconductors with  critical temperatures $T_c=8.6$~K and %%%
3.4~K, respectively.\cite{hulliger,ts} $\text{Nb}_5\text{Te}_4$
and $\text{Ti}_5\text{Te}_4$ were reported to behave as normal
metals down to 1.1~K.\cite{hulliger} Recently, a monoclinic form
of $\text{Nb}_{4.7}\text{Te}_4$ has been synthesized using a
chemical transport reaction.\cite{dron92}

In this work we found that $\text{Nb}_{5-\delta}\text{Te}_4$ is a
bulk superconductor with $T_c=$0.6--0.9~K. In the normal state
quantum interference effects determine the electronic transport
properties in a broad temperature
range. The observed nonmonotonic behavior of the %%%
temperature dependence of the resistance and the low-temperature
positive magnetoresistance are interpreted in terms of weak
localization and electron-electron interaction effects in
disordered conductors. The temperature dependence of the dephasing
time $\tau_\varphi$ is determined by electron-phonon interaction.
Saturation of the dephasing time at low temperatures is ascribed
to scattering on magnetic impurities.

\section{Experimental}
$\text{Nb}_{5-\delta}\text{Te}_4$ single crystals were prepared by chemical
vapor transport from powders of the elements Nb (Johnson Matthey Inc. 99.99\%
metals basis, excluding Ta, Ta $<$ 500 ppm) and Te (Johnson Matthey Inc.
99.99\%) in evacuated silica tubes with I$_2$ used as a transport agent.
The samples were annealed for 1 day at 750$^{\circ}$C, %%%followed
the following 30 days at 980$^{\circ}$C, and then were slowly cooled
to room temperature. X-ray powder-diffraction patterns
were collected with a Stoe  diffractometer using %%%
(Cu $K_{\alpha 1}$ radiation). They show a body-centered
tetragonal unit cell with lattice parameters $a=10.234(1)$~{\AA}
and $c=3.7021(6)$~{\AA} which are in agreement with those observed
by Selte and Kjekshus. \cite{set63} Energy dispersive x-ray
analysis carried out on the three crystals used for the resistance
measurements
reveals a Nb deficiency of %%%of
$\delta=0.23(4)$. A single crystal x-ray diffraction measurement
with a Stoe image plate detector system  was carried out on a
small crystal. It shows that the Nb deficiency is associated with
the outer site of the Nb octahedral chains (Wyckoff position
8$h$). Single crystal x-ray measurements performed  on the three
crystals used for the resistance measurements gave lattice
parameters in good agreement with the powder diffraction data.

The heat capacity of a 6.3~mg crystal was measured in a Quantum Design PPMS
relaxation calorimeter in the temperature range 0.3--5~K and external fields of
0~T and 9~T. The sample was attached with a minute amount %%%Mis see t"ahendab?
of Apiezon vacuum grease to the calorimeter platform the heat capacity
of which was determined in a separate run and subtracted %%%
afterwards.

For resistance measurements %%%
needle-like crystals with the needle axis collinear with the crystallographic
$c$-axis were selected. Crystals were 3--5~mm long with cross section of
0.005--0.05~mm$^2$. Four electrical contacts were placed along the needle at
distances of $\sim$2~mm with the two outer contacts as the current contacts
such that the electrical current was directed along the $c$-axis.

Crystals stored in air get covered by a high-resistive oxide
layer.  Low-resistance ohmic contacts can be achieved after this
layer has been etched off with an Ar plasma in a vacuum chamber
followed by the immediate deposition of the gold contact pads
through a shadow mask. Similar good contacts can also be made by
gluing gold wires with silver epoxy resin on a crystal surface
freshly cleaved in  an argon atmosphere. The results do not depend
on the way contacts have been applied.

The resistance was measured by a dc four-probe technique using a
high resolution nanovoltmeter (7\,1/2 digits) and a Keithley 2400
current source. Measurements were performed using a variable
temperature Oxford $^4$He cryostat with a superconducting magnet.
The rotatable sample holder with the rotation axis perpendicular
to the magnetic field allows us to align the crystal either
perpendicular or nearly parallel to the magnetic field.
Magnetoresistance was measured at constant temperature in fields
up to 11~T. The superconducting transition and its dependence on
the magnetic field was measured in a %%%
home-built single shot $^3$He refrigerator. The bias currents were
chosen such that the  power dissipated in the sample remained
below 30~pW and 2~$\mu$W for the measurements in the $^3$He and
the $^4$He measurements systems, respectively.

\section{RESULTS AND DISCUSSION}
\subsection{Superconductivity}

\begin{figure}
\includegraphics[width=8cm]{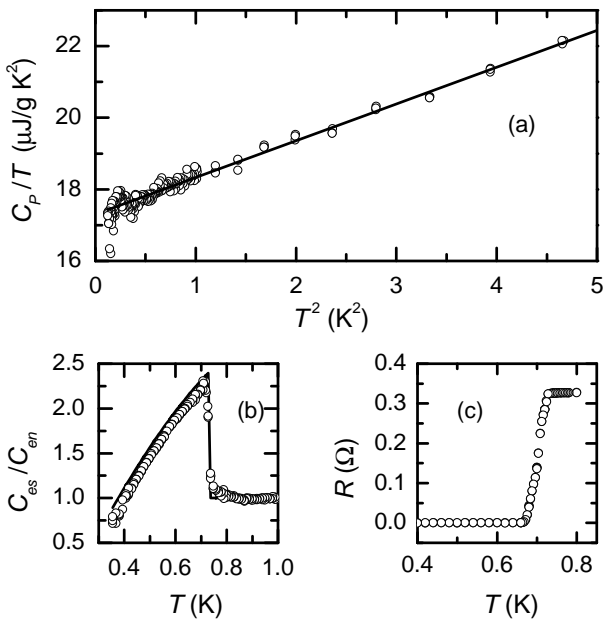}
\caption{\label{CT} (a) Specific heat $C_p$  of
$\text{Nb}_{5-\delta}\text{Te}_4$ measured in a magnetic field of
9~T plotted as $C_p/T$ vs $T^2$. The symbols represent the
experimental data and the solid line a fit to the equation
$C_p=\gamma \,T+\beta \, T^3$  with $\gamma=$17.29~$\mu$J/g\,K$^2$
and $\beta=1.027$~$\mu$J/g\,K$^4$. (b) The ratio of the electronic
heat capacity in the superconducting state $C_{es}$ to that in the
normal state $C_{en}$ as a function of temperature. The solid line
is the heat capacity anomaly expected for a BCS superconductor
with $T_c=0.73$~K. (c) The superconducting transition of Nb09 (see
TABLE~\ref{tab}) measured from the electrical resistance at 0~T.}
\end{figure}

A total of six crystals was checked for superconductivity by
resistance measurements. All investigated samples show a
transition to superconductivity with critical temperatures in the
range 0.60--0.88~K and transition widths between 0.012--0.18~K. A
typical resistive transition is shown in Fig.~\ref{CT}(c).
Measurements in magnetic fields reveal a linear dependence of the
upper critical field $H_{c2}(0)$ as a function of temperature down
to 0.35~K, with a slope of $d H_{c2}/d T=-1.2$~T/K.

The heat capacity measured in zero field is characterized by an
anomaly typical for a transition to
superconductivity %%%
(Fig.~\ref{CT}(b)). The anomaly disappears in a field of %%%
9~T. In the range 0.3--2~K the heat
capacity %%%
can be well described by a polynomial %%%
$C_p=\gamma T+\beta T^3$, %%%
where the linear and cubic terms are electronic and lattice
contributions, respectively (Fig.~\ref{CT}(a)). %%%
The fit yields the Sommerfeld term $\gamma=$17.29~$\mu$J/g\,K$^2$,
and the phonon term $\beta=1.027$~$\mu$J/g\,K$^4$ corresponding to
the Debye temperature $\Theta_{D}=$259~K. The superconducting
anomaly
within experimental %%%
errors agrees well with the anomaly expected for a BCS-type
superconductor. Resistivity and heat capacity results clearly
prove that $\text{Nb}_{5-\delta}\text{Te}_4$ is a bulk
superconductor.

The electronic density of states at the Fermi level $\nu$ was calculated from
the Sommerfeld term using the formula \cite{mcmillan} $ \gamma = \pi ^2 \nu
k_B^2 (1+ \lambda )/3$, where $k_B$ is the Boltzmann constant. The
electron-phonon coupling constant $ \lambda=0.44$ was estimated from the
McMillan empirical formula \cite{ mcmillan}
\begin{equation}
\lambda=\frac{1.04+\mu^\ast \ln \left(\Theta_D/1.45
T_c\right)}{(1-0.62\mu^\ast)\ln\left(\Theta_D/1.45
 T_c\right)-1.04},
\end{equation}
with the Coulomb pseudopotential %%%of
$\mu^\ast=0.13$, %%%,
which is typical for the transition metals. In
$\text{Nb}_{5-\delta}\text{Te}_4$ we obtain $\nu=1.59 \times
10^{47} \text{J}^{-1}\,\text{m}^{-3}$, which is comparable to the
density of states in good metals such as silver, copper, and gold.

\begin{table*}
\caption{\label{tab}Values of relevant parameters for
$\text{Nb}_{5-\delta}\text{Te}_4$ samples. $T_c$ is the superconducting
temperature (midpoint of resistive transition), $D$ is the diffusion constant %%%
determined from $H_{c2}(T)$, and $\rho_0^*$ is the resistivity measured at
300~K. $\rho_0$, $c_{\text{MT}}$ , and $k_C$ are fitting parameters of the
magnetoresistance; $K$, $n$, and $\tau_0^{-1}$ describe the temperature
dependence of the dephasing time; $k_C^*$, $L$, $p$, and the electron screening
parameter $\tilde{F}$ were determined from the temperature dependence of the
resistance.}
\begin{ruledtabular} \begin{tabular}{cccccccccccccc}
Sample&$T_c$&$D$&$\rho_0^*$&$\rho_0$&$c_{\text{MT}}$&$k_C$&$k_C^*$&$K$&$n$&$
\tau_0^{-1}$&$\tilde{F}$&$L$&$p$\\
&(K)&(cm$^2$/s)&$(\mu\Omega\,\text{cm})$&$(\mu\Omega\,\text{cm})$&&&&(s$^{-1}$
\,K$^{-n}$)&&(s$^{-1}$)&\\
\hline \\ Nb10 & 0.79 & 0.924 & 241 & 249 & 0.768 & 0.379 &
0.371&$7.75\times10^8$&2.45&$1.77\times10^{10}$&$\;$0.048&$\;90\times10^{-6}$&1.28\\
Nb19 & 0.88 & 0.883 & 231 & 254 & 0.730 & 0.668 &
0.392&$4.39\times10^8$&2.68&$0.58\times10^{10}$&-0.323&$\;71\times10^{-6}$&1.37\\
Nb09 & 0.70 & 0.890 & 304 & 294 & 0.769 & 0.348 &
0.314&$9.52\times10^8$&2.42&$2.61\times10^{10}$&-0.234&$129\times10^{-6}$&1.19\\
\end{tabular} \end{ruledtabular} \end{table*}

\subsection{Electrical Resistivity}
The temperature dependence of the resistivity was measured on six
crystals. All samples show resistivities between 200--300 $\mu
\Omega$\,cm (see Table~\ref{tab}) and a weak
variation with temperature %%%,
(less than 2\% in the range of 2--270~K, see Fig.~\ref{RTFig}).
This finding points to  strong scattering of conduction electrons.
This is  also reflected in the
electron diffusion constant $D$ %%%being
($\approx$ 1~cm$^2$/s) and %%%
a low Hall mobility.
The diffusion constant was determined %%%via
from the temperature dependence of the critical magnetic field
using $D=-4 k_B/[\pi e \left( d H_{c2}/d T \right)]$, where $e$ is
the electron charge.\cite{gennes} An alternative estimation of $D$
from the resistivity $\rho_0$ using the Einstein relation
$1/\rho_0=e^2 D \nu$ gives $D=0.9$--1.2~cm$^2$/s, which is
consistent with the value determined from $ H_{c2}(T)$ to within
20\% (see Table I). The agreement is reasonable considering
uncertainties in $\rho_0$ and $\nu$. As being more reliable, the
value of $D$ estimated from the critical magnetic field
measurements  will be used in the analyses below. The Hall
constant is negative indicating that electrons contribute to the
charge transport. The Hall mobility is low, 0.25 cm$^2$/(V\,s), which %%%
in the free electron model gives %%%an
an electron mean free path of
 $\ell \simeq 2~{\text{\AA}}$ of the order of interatomic
distances. The origin of strong electron scattering in
$\text{Nb}_{5-\delta}\text{Te}_4$ is
%%%opened.
an open question. The Nb deficit is definitely essential, however
deformations of the low-dimensional structure may also contribute
to the scattering.

A closer inspection of the temperature dependence of the
resistance reveals qualitatively different behavior above
$\sim$~50~K with either positive or negative temperature
coefficients (see Figure~\ref{RTFig}(a)). The magnetoresistance
behavior and the temperature dependence of the resistance of three
representative samples Nb10, Nb19, and Nb09 have been studied in
detail and the analysis is described in the following.

\subsubsection{Magnetoresistance}
\begin{figure}
\includegraphics[width=8cm]{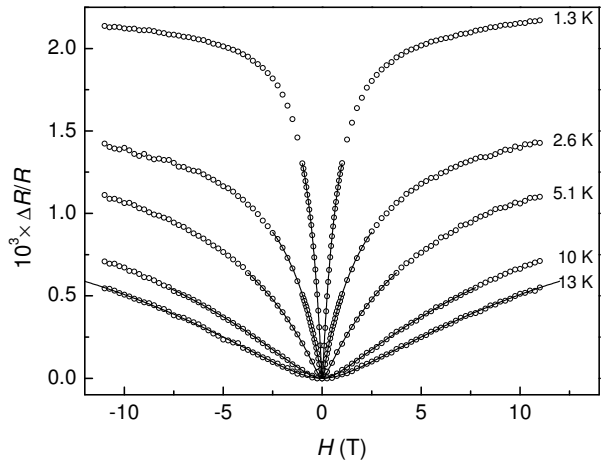}
\caption{\label{RHfig} The normalized magnetoresistance $\Delta
R/R=[R(H)-R(0)]/R(0)$ in Nb19 (see TABLE~\ref{tab}) versus the
magnetic field
for several temperatures. The symbols %%%presents
are the experimental data and the lines are the fits with
Eq.~(\ref{RH}) for the three-dimensional quantum interference
corrections.}
\end{figure}

Figure~\ref{RHfig} shows a typical dependence of the resistance on
the magnetic field measured for temperatures between 1.3~K and
13~K. The small positive magnetoresistance ($<0.3\%$) with a
minimum centered at zero magnetic field, which broadens as the
temperature is increased, is a typical fingerprint of quantum
interference effects of the conduction electrons.\cite{alt85} The
positive magnetoresistance is expected in disordered
superconductors containing heavy elements in which both the
scattering on virtual Cooper pairs and weak antilocalization of
conduction electrons induced by spin-orbit scattering are
essential. We found that the magnetoresistance is independent of
the orientation of the magnetic field. This result is rather
unexpected in view of the anisotropic crystal structure. We
tentatively ascribe this finding to averaging effects caused by
strong electron scattering.

In the following we compare our magnetoresistance data with results of a theory
treating three-dimensional quantum interference corrections to resistivity.  In
the limit of low fields the relative change of the resistance is expressed by
the equation \cite{kawa,alt81}
\begin{equation}
\frac{R(H)-R(0)}{R(0)} =A \frac{e^2}{2 \pi^2 \hbar} \sqrt{\frac{e H}{\hbar}}
f_3 \left( \frac{4 e D \tau_\varphi H}{\hbar} \right) + B H^2 \label{RH},
\end{equation}
where $\hbar$ is the Planck constant and $f_3
(1/x)=2(\sqrt{2+x}-\sqrt{x})-[(0.5+x)^{-1/2}+(1.5+x)^{-1/2}]+(2.
03+x)^{-3/2}/48$.\cite{bax} In %%%
Eq.~(\ref{RH}), the first term with %%%
the prefactor $A=\rho_0 / 2$ describes weak antilocalization in the limit of
strong spin-orbit scattering, $\tau_\varphi^{-1} \ll \tau_{so}^{-1}$, where
$\tau_{so}^{-1}$ is the spin-orbit scattering rate. In superconductors
for $T > T_c$ %%%
the scattering on virtual Cooper pairs, the Maki-Thompson-Larkin effect, is
also described by the first term
in %%%
Eq.~(\ref{RH}) with a temperature-dependent prefactor $A=\rho_0 c_{\text{MT}}
\beta (T/T_c)$. Here $\beta$ is the function tabulated in
Ref.~\onlinecite{lar80} and $c_{\text{MT}}=1$ and 0.25 in the limits of weak
and strong spin-orbit
scattering, respectively. \cite{alt81} This expression is valid for %%%
magnetic fields $H \ll k_B T/eD$. Expressions %%%extending this limit
for larger fields have been derived for %%%
the two-dimensional %%%systems
case. \cite{lopes,brenig} For the three-dimensional case, results
for an extended range of fields have only been analyzed
numerically for an Mg$_{67}$Zn$_{33}$ alloy sample.\cite{gey}

In the field limit set by the Maki-Thompson-Larkin effect, $H \ll
k_B T/eD$, contributions of the classical magnetoresistance and
electron-electron interaction are described by the quadratic term
in Eq.~(\ref{RH}). The magnetoresistance due to the
electron-electron interaction in the Cooper channel is negative
and diverges as $T$ approaches $T_c$. Its contribution is
described by the parameter \cite{alt81}
\begin{equation}
B=-8.49 \times 10^{-3}\left(\frac{D}{\hbar k_B T}\right)^{3/2}\frac{e^4\rho_0
k_C}{\ln(T/T_c)} \label{BT}, \label{cooper}
\end{equation}
with $k_C=1$ and 0.25 in the limit of weak and strong spin-orbit
scattering, respectively. In %%%
fields $H\ll k_B T/g\mu_B$   the electron-electron interaction in the
diffusion channel is described by the coefficient
\begin{equation}
B=9.5 \times 10^{-4}\rho_0 \tilde{F}\sqrt{\frac{k_B T}{\hbar D}}\left(\frac{g
\mu_B}{k_B T}\right)^{2},
\end{equation}
where $\mu_B$ is the Bohr magneton and $g$ is the gyromagnetic
ratio.\cite{lee85} The electron-screening parameter $\tilde{F}$ approaches 1 in
the limit of complete screening and 0 if screening is negligible. In
superconductors due to the exchange with virtual phonons, negative values of
$\tilde{F}$ are expected.\cite{alt85, lin96}.

\begin{figure}
\includegraphics[width=8cm]{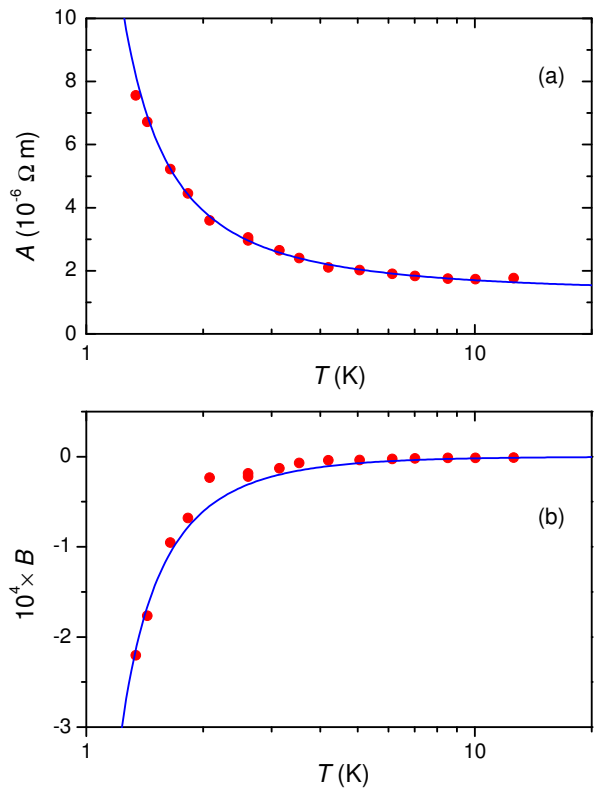}
\caption{\label{ABfig} Temperature dependence of the coefficient $A$ %%%
(the upper panel) and $B$ %%%
(the lower panel) for sample Nb19 (see TABLE~\ref{tab}).
Solid lines are the best fit %%%of
for $A$ and $B$ with Eq.~(\ref{AT})
and %%%Eq.~
(\ref{BT}), respectively.}
\end{figure}

\begin{figure}
\includegraphics[width=8cm]{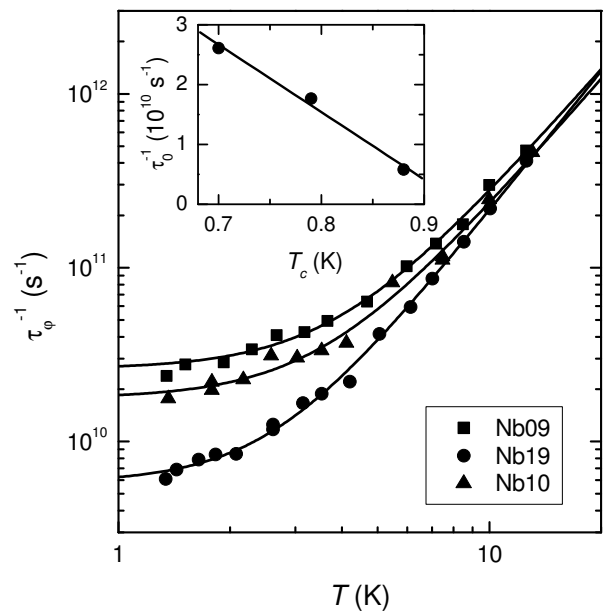}
\caption{\label{tauFig} The dephasing rate $\tau_\varphi^{-1}$ %%%
vs.\ temperature for samples Nb09, Nb10, and Nb19 (see
TABLE~\ref{tab}). Symbols represent experimental data and the
solid lines are the best fit with Eq.~(\ref{tau}). Inset:
$\tau_0^{-1}$ dependence on the superconducting transition
temperature.}
\end{figure}

In the magnetic field range $|H|\leq0.5k_BT/eD$ the magnetoresistance data were
fitted to Eq.~(\ref{RH})  using $A$, $B$, and $\tau_\varphi$  as fitting
parameters. The fits shown as solid lines in Fig.~\ref{RHfig} are in good
agreement with the experimental data for all temperatures. The temperature
dependence of  $A$, $B$, and
$\tau_\varphi$ %%%,
is %%%
displayed in Fig.~\ref{ABfig} and \ref{tauFig}. $A$ is positive
and increases as the temperature decreases below 9~K. The
temperature dependence of $A$ is well described by the formula
\begin{equation}
A=\rho_0\left[c_{\text{MT}}\beta(T/T_c)+ 1/2 \right], \label{AT}
\end{equation}
with the Maki-Thompson-Larkin correction and weak antilocalization due to
strong spin-orbit scattering as first and
second %%%
terms, respectively. The  fit parameters $\rho_0$ coincide within 10\% with the
measured values of the electrical resistivities (see
Table~\ref{tab}). %%%In the rest of this paper
Below, the fitted values for $\rho_0$ will be used for the description of the
temperature dependence and the electron-electron interaction in the Cooper
channel. For strong spin-orbit scattering, $c_{\text{MT}}$ is expected to be
0.25 (see above). The fits rather give  $0.730 < c_{\text{MT}} < 0.769$
indicating that the strong spin-orbit scattering limit for the
Maki-Thompson-Larkin correction is not realized.
Such a situation has often been observed in %%%other
two- and three-dimensional systems.\cite{ger83,lin94}

$B$ is negative, and its characteristic temperature
dependence (see Fig.~\ref{ABfig}(b)) allows %%%
us to identify the origin of the second term in Eq.~(\ref{RH}) with the
electron-electron interaction in the Cooper
channel. The %%%
single-parameter fit %%%to
with Eq.~(\ref{cooper}) is in %%%a
good agreement with the experimental data. As in the case of the
Maki-Thompson-Larkin correction, the value of the fitting parameter $k_C$ (see
Table~\ref{tab}) lies between the strong and weak spin-orbit scattering limits.

Figure~\ref{tauFig} shows the temperature dependence of $\tau_\varphi$ for
samples Nb09, Nb10, and Nb19. The solid lines correspond to fits with the
formula
\begin{equation}
\tau_\varphi^{-1}=K T^n + \tau_0^{-1}, \label{tau}
\end{equation}
with the fitting parameters $\tau_0^{-1}$, $K$, and $n$  listed in
Table~\ref{tab}. In Eq.~(\ref{tau}) the exponent  $n\approx 2.5$ indicates that
the temperature dependence of $\tau_\varphi^{-1}$ arises from electron-phonon
interaction. A similar temperature dependence with comparable exponents has
been observed for NbC, Sb, and
$\text{Pd}_{60}\text{Ag}_{40}$.\cite{NbC,lin00,lin02b} Dephasing due to
electron-electron interaction with %%%
small energy transfer is also described by a power law but with a
smaller exponent, $n=3/2$, and the coefficient\cite{alt79,alt81ss}
\begin{equation}
K=\frac{k_B^{3/2}}{12\sqrt{2}\pi^3\hbar^{5/2}\nu D^{3/2}}.
\end{equation}
%%%
Numerical estimates with this formula give a dephasing rate which is more than
two orders of magnitude smaller than that observed in the experiment.
Therefore, as is often the case in three-dimensional systems, this mechanism
can be neglected for $\text{Nb}_{5-\delta}\text{Te}_4$.\cite{bird}

Deviations of $\tau_\varphi^{-1}(T)$ from the power law are
observed below 5~K and are described by the parameter $\tau_0
^{-1}$. The fitted values for $\tau_0 ^{-1}$ correlate with the
critical temperatures (see inset in Fig.~\ref{tauFig}). This
points to scattering on magnetic impurities which
decreases both $T_c$ and the dephasing time. %%%
The spin-flip scattering rate $\tau_s ^{-1}$ is proportional to
the concentration of magnetic impurities $x_{\text{mag}}$
according to\cite{map}
\begin{equation}
\frac{1}{\tau_s}=\frac{x_{\text{mag}}}{\pi\hbar\nu}\frac{\pi^2S(S+1)}
{\pi^2S(S+1)+\ln^2(T/T_K)},
\end{equation}
where $S$ is the impurity spin and $T_K$ is the Kondo temperature.
For impurities with $S>3/2$ and $T_K <0.3$~K we obtain
$x_{\text{mag}} =9\times 10^{-6}$, $30\times10^{-6}$, and
$42\times 10^{-6}$ per host atom for Nb19, Nb10, and Nb09
respectively. These concentrations correlate with the purity of
the starting materials used for the sample preparation. Additionally, we
carried out temperature dependent magnetization measurements on
$\text{Nb}_{5-\delta}\text{Te}_4$. These reveal a Curie type
susceptibility contribution with a Curie constant which is
consistent with these impurity estimates if we assume Fe$^{3+}$
ions ($S=5/2$) as major impurities.

The effect of the magnetic impurities on $T_c$ is characterized by
the slope $d T_c/d x _{\text{mag}}$. In
$\text{Nb}_{5-\delta}\text{Te}_4$ we have $d T_c/d
x_{mag}=-5.3\times10^3$, which is of the same order of magnitude
as observed for Zn-Ni, Zn-Co, and Al-Mn alloys.\cite{boate}

Other mechanisms of the dephasing saturation can be excluded %%%
based on experimental data. %%%
Electron scattering by the exchange of superconducting fluctuations modifies
the temperature dependence of $\tau_\varphi$ near $T_c$. In
various systems this effect %%%appears
manifests itself either through a negative temperature
coefficient\cite{gordon, lin94, lin98, stol} or a
saturation\cite{lin98} of $\tau_\varphi (T)$. In the latter case
an increase %%%of
in $\tau_0^{-1}$ with increasing $T_c$ is expected %%%
which is not seen in the experiments. This result is also in line with the
conclusion that contribution of the electron-electron interaction in
$\tau_\varphi$ is negligible.

Nonequilibrium effects of the conduction electrons can lead to
saturation of $\tau_\varphi$ in several cases.\cite{bird,ova01}
Measurements of $R$ and $\tau_\varphi$ at different bias currents
show that heating effects are not
important in our experiment. %%%
Besides, the internal thermometers %%%????
associated with the Maki-Thompson-Larkin effect and the electron-electron
interaction in the Cooper channel (Fig.~\ref{ABfig}) follow the
variation of %%%
the external temperature, while $\tau_\varphi$ %%%is saturating
saturates. This also indicates that for the chosen measuring parameters the
system is in thermal equilibrium.

We note that our $\tau_0$ values agree with the scaling relation
for this parameter found by Lin and Kao for numerous
three-dimensional polycrystalline alloys with small diffusion
constant ($D$=0.1--10~cm$^2$/s).\cite{lin01} However, in view of
the magnetic impurities found in our samples the agreement rather
appears to be accidental. Investigations crystals with a reduced
impurity level are necessary to compare the saturation behavior of
$\tau_\varphi$ in single- and polycrystalline materials.

\subsubsection{Temperature dependence of resistance}
\begin{figure*}
\includegraphics[width=16cm]{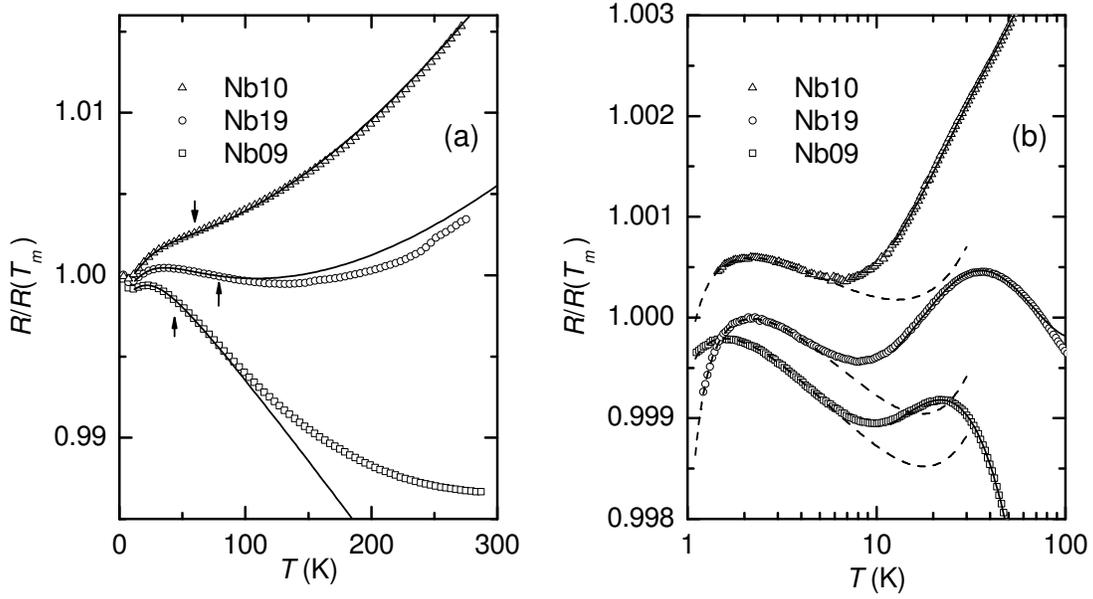}
\caption{\label{RTFig} The resistance %%%
vs.\ temperature in the linear %%%
(the left panel) and in the logarithmic %%%
(the right panel) temperature scales for samples Nb09, Nb10, and
Nb19 (see TABLE~\ref{tab}). The resistances are normalized to the
value at the low-temperature maximum $T_m$. For better visibility
curves and symbols of samples Nb09
and Nb10 are shifted vertically. The dashed lines are the best %%%
fits of
the data in the %%%interval
range 1.2--4.2~K %%%to
with Eq.~(\ref{rt}). The solid lines are the best %%%
fits of the data %%%to
with Eq.~(\ref{rt2}) in the %%%interval
range from 12 K to the value shown %%%as an
by arrows in left panel.}
\end{figure*}

Figure~\ref{RTFig} displays the temperature dependence of the
resistance for $T>T_c$. Up to 270~K,
the resistance varies by less than 2\% %%%,
which indicates that the %%%
contribution of the electron-phonon scattering to the resistivity
is largely suppressed. A closer inspection reveals that the
temperature dependence is nonmonotonic and characterized by one
maximum around 2~K and a second maximum/hump between 20~K and
40~K. We ascribe this behavior to an interplay of weak
localization and electron-electron interaction effects, some of
which also contribute to the magnetoresistance. At
$T>50$~K the temperature dependence is %%%
sample dependent. Typical behaviors are illustrated by samples
Nb09, Nb10, and Nb19 in Fig.~\ref{RTFig}(a). While in Nb09 a
negative temperature coefficient extends up to  room temperature,
Nb10 shows a metallic behavior with a positive temperature
coefficient at all temperatures. Nb19 with a resistance minimum at
130~K falls between these two limiting cases.

%%%

We begin the data analysis with the %%%
low-temperature region around the
first maximum. We associate it %%%to
with the interplay of the superconducting fluctuation effects and the
electron-electron interaction in the diffusion channel. %%%At
For these temperatures the weak antilocalization is suppressed because
of %%%
the saturation of $\tau_\varphi$. In the range 1.2--4.2~K the experimental data
are well fitted by the formula\cite{alt85,kawa}
\begin{eqnarray}
\frac{R(T)}{R(T_m)}=\frac{e^2}{2\pi^2\hbar}\rho_0 \Bigg\{ \sqrt{\frac{k_B
T}{\hbar D}} \Bigg[-F_d
- \frac{0.915 k_c}{\ln(T/T_c)} \nonumber \\
- \frac{4}{3\pi^2} c_{\text{MT}} I_{\text{MT}}(T/T_c,\tau_\varphi)\Bigg]+\sqrt
{\frac{1}{4\tau_\varphi (T)D}} \,\Bigg\}+R_0. \label{rt}
\end{eqnarray}
The first term in Eq.~(\ref{rt}) corresponds to the electron-electron
interaction in the diffusion channel with $F_d=0.915\,(2/3-3\tilde{F}/4)$. The
electron-electron interaction in the Cooper channel, %%%
the Maki-Thompson-Larkin correction, %%%(for the temperature dependence
%%%this correction is called Maki-Thompson correction)
and the weak antilocalization, also observed in the magnetoresistance, are
described
by the second, third and fourth %%%
terms, respectively. $I_{\text{MT}}$ is a triple integral %%%
(see Eq.~(56) in Ref.~\onlinecite{rei92}), which determines the
dependence of the Maki-Thompson-Larkin correction on $T/T_c$ and
$\tau_\varphi$. The values of the fitting parameters
$\tilde{F}$, $k_C$ are listed in Table~\ref{tab}. %%%
The values of $k_C$ determined from $R(T)$ and $R(H)$ are consistent with each
other. As expected for superconductors, $\tilde{F}$
is negative in Nb09 and %%%
Nb19.  Their values -0.234 and -0.323 for Nb09 and Nb19,
respectively, are close to %%%
those found, for example, in $\text{Ti}_{1-x}\text{Sn}_x$ and
$\text{Ti}_{1-x}\text{Ge}_x$ alloys %%%,
which show higher critical temperatures.\cite{lin96} In
Nb10 $\tilde{F}$ is larger and positive, $\tilde{F} =0.048$. We
attribute this to the
Boltzmann transport term %%%
which extends to lower temperatures in this sample and %%%
is not included in Eq.~(\ref{rt}). %%%
Indeed, the sample Nb10 exhibits the best metallic properties and
consequently a larger contribution of the Boltzmann term is expected.

As the temperature is increased above 13~K %%%
the weak antilocalization contribution starts to dominate. This results in a
positive slope of $R(T)$ (see the
extrapolation curve %%%of
from Eq.~(\ref{rt}) in Fig.~\ref{RTFig}(b)). However, %%%compared to
in comparison with the experiment the calculated transition is shifted to
somewhat higher
temperatures. This discrepancy indicates that in %%%that
the considered range a mechanism with positive $R(T)$ slope, %%%
e.g.\ the Boltzmann term, is essential in all samples.

%%%
We associate the high-temperature maximum in $R(T)$ %%%
(a "bulge" in Nb10) with %%%
the emergence of the weak localization %%%as
with increasing  temperature so that $\tau_\varphi^{-1}\gg\tau_{so}^{-1}$. In
this case
the temperature dependence of the resistance %%%(around the second maximum
%%%describe)
is %%%approximated
described by %%%with
the formula
\begin{widetext}
\begin{equation}
\frac{R(T)}{R(T_m)}=\frac{e^2}{2\pi^2\hbar}\rho_0\left[-F_d \sqrt{\frac{k_B
T}{\hbar D}} -3\sqrt {\frac{1}{4\tau_{so}D}+\frac{1}{4\tau_\varphi (T)D}}
+\sqrt {\frac{1}{4\tau_\varphi (T)D}} \;\right]+LT^{p}+R_0, \label{rt2}
\end{equation}
\end{widetext}
where the extrapolation of the low-temperature behavior of the
electron-electron
interaction in the diffusion channel %%%
(the first term) and %%%
the weak localization and antilocalization \cite{fuk81} %%%
(the second and third terms) is used. The Boltzmann transport %%%
term is described by the term $L T^{p}$  and the superconducting
fluctuation effects are ignored. The solid lines in Fig.~\ref{RTFig} represent
the
best fit %%%to
with Eq.~(\ref{rt2}) with parameters $L$, $p$, $R_0$, and
a universal $\tau_{so}^{-1}=2.3\times10^{12}$~s$^{-1}$ %%%,
for all samples. %%%One can see
It can be seen that Eq.~(\ref{rt2}) %%%approximate
gives a good approximation %%% acceptably not only
for the feature around 20--40~K %%%but also
and extrapolates %%%the temperature dependence
to higher temperatures reproducing both "metallic" and "insulating" behavior.
In all cases the
temperature exponent of the Boltzmann term lies in %%%a
the range %%%of
$p=$1.2--1.4 (see Table~\ref{tab}). This finding points to
non-Fermi-liquid behavior of conduction electrons in
$\text{Nb}_{5-\delta}\text{Te}_4$. Usually the quantum
interference of conduction electrons in disordered conductors is
considered as a low-temperature effect. Its extension to higher
temperatures is possible when the dephasing length is still larger
than the electron mean free path, $\sqrt{D \tau_\varphi} \gg\ell$.
This occurs in systems with strong potential scattering such as
Ti-Al alloys \cite{lin93}, Mo/Si multilayers \cite{MoSi}, and
ion-implanted polymers \cite{poly}, where the interference effects
persist above 250~K. Therefore in
$\text{Nb}_{5-\delta}\text{Te}_4$, which is also characterized by
a short mean free path, high-temperature quantum interference of
conduction electrons is also possible.

In summary, we studied %%%
the dependence of the resistance on the magnetic field and temperature in
$\text{Nb}_{5-\delta}\text{Te}_4$ single crystals with
$\delta=$0.23. The compound is %%%
a superconductor with $T_c=$0.7--0.9~K. Both the magnetoresistance and the
temperature
dependence of the resistivity are quantitatively well described in the %%%
framework of the theory of quantum interference effects in disordered
conductors. %%%
The electron dephasing times extracted from the magnetoresistance
are determined by the electron-phonon interaction and scattering
on magnetic impurities.

\begin{acknowledgments}
The authors thank V.~Duppel for the microprobe analysis,
E.~Br\"ucher for magnetic susceptibility measurements, and
D.~Martien for assisting the heat capacity investigations. This
work was supported by the Estonian Science Foundation Grant
No.~5033. ASt kindly acknowledges support from the
Max-Planck-Gesellschaft. %%% for a Forschungsstipendium.
\end{acknowledgments}
% Add \usepackage{longtable} and the longtable (or longtable*}
% environment for nicely formatted long tables. Or use the the [H]
% placement option to break a long table (with less control than
% \begin{table}%[H] add [H] placement to break table across pages
% \begin{ruledtabular}
% Lines of table here ending with \\
% \end{tabular}
% \end{ruledtabular}
% \end{table}

% Surround table environment with turnpage environment for landscape
% table
% \begin{table}
% \begin{ruledtabular}
% \end{tabular}
% \end{ruledtabular}
% \end{table}

\bibliography{art}

\end{document}